# Quantum interference between autonomous dissimilar quantum light sources for hybrid quantum networks

Kyu-Young Kim[1†], Heewoo Kim[2†], Dong Hyun Park[1], Jinhyuk Bea[2], Gyeongmin Ju[2], Suk In Park[3], Jin Dong Song[3], Je-Hyung Kim[1,4‡] and Han Seb Moon[2,4]*

[1]Department of Physics, Ulsan National Institute of Science and Technology; Ulsan, 44919, Republic of Korea

[2]Department of Physics, Pusan National University; Busan, 46241, Republic of Korea

[3]Center for Opto-Electronic Materials and Devices Research, Korea Institute of Science and Technology; Seoul, 02792, Republic of Korea

[4]Quantum Sensors Research Center, Pusan National University; Busan, 46241, Republic of Korea

Corresponding authors. Email: [‡]jehyungkim@unist.ac.kr and [*]hsmoon@pusan.ac.kr

**Abstract:**
Hybrid quantum systems play a crucial role in advancing scalable and versatile quantum networks as they combine the strengths of different quantum platforms. An important challenge for the development of hybrid quantum networks lies in interfacing heterogeneous quantum nodes and distributing entanglement among them. Single photons emitted from these dissimilar quantum nodes typically show distinct spectral and temporal properties. Therefore, they necessitate spectral filtering and temporal synchronization, which introduce significant photon losses and require additional resources. In this work, we successfully generate indistinguishable photons from two distinct quantum systems of a warm atomic ensemble and a solid-state quantum dot. Remarkably, quantum interference between dissimilar sources is achieved without additional spectral filtering and time synchronization, which enables autonomous quantum nodes for a hybrid quantum network. $^{133}$Cs atomic ensemble can efficiently generate heralded single photons at the wavelength of 917 nm of the $6P_{3/2} - 6D_{5/2}$ transition, while the single photons emitted from an InAs/GaAs quantum dot can be tuned to match the $^{133}$Cs transition wavelength. Our dense warm atomic ensemble and cavity-coupled quantum dot can efficiently generate bright and resonant single photons at detection rates approaching MHz, respectively. More importantly, these single photons exhibit inherent spectral similarities not only in the wavelength but also in the spectral linewidth, achieving a high spectral overlap of 0.92. Such intrinsic compatibility between dissimilar quantum sources is essential to leverage the advantages of different quantum platforms, paving the way toward a large-scale and functional hybrid quantum network.

# Introduction

Quantum networks consist of quantum nodes and quantum channels, which are based on quantum memories and quantum light sources [1-3]. Implementing highly entangled photon states and entanglement swapping are also essential parts for correcting errors and extending the communication distance [4,5]. Achieving the source of coherent single photons that have narrow linewidth, high brightness, spectral uniformity, and compatibility with quantum memories is a critical requirement for a large-scale quantum network. A variety of quantum platforms, such as spontaneous parametric down conversion (SPDC) [6,7], atoms [3,8,9], color centers [1,2,5], and quantum dots (QDs) [10-13], have been proposed and experimentally demonstrated quantum key distributions, quantum teleportation, and entanglement distribution. Despite these advances, realizing a scalable and functional quantum network remains challenging due to the inherent limitations of each quantum platform, such as photon loss, limited coherence times, probabilistic operation, and integration difficulties.

Hybrid quantum architectures that combine different types of quantum light sources can address these challenges by leveraging the advantages of each platform. For example, QDs can serve as on-demand single photons with high brightness, high purity, and near Fourier-transform-limited linewidth [14-16]. These solid-state emitters are particularly suited for integration with nanophotonic structures for tailoring light-matter interactions [17,18], as well as optical fibers [19] or photonic-integrated chips [20,21], enabling scalable miniaturization [22]. However, QDs suffer from the spectral randomness and difficulties in storing photons, which limit long-distance interactions between remote QDs. In contrast, atomic systems with their naturally identical energy levels ensure uniform interactions across quantum nodes [3,8,23] and offer reliable frequency standards within a quantum network. This will allow the distribution of tasks in a quantum network such that QDs handle high-rate photon generation for fast transmission, while atoms manage storage and synchronization of distant QD emissions [24-26].

To realize quantum interference and to physically implement such a modular quantum architecture, achieving high identity of emissions from each source in both central frequency and physical modes, such as spectral and temporal profiles, is essential. Previously, two-photon interference between dissimilar sources has been demonstrated from various platforms, including sunlight (thermal light)-QD [27,28], laser (coherent light)-QD [29,30], and SPDC (heralded single photons)-QD [31,32]. However, these approaches require spectral filters to compensate for large mismatches in spectral linewidth between the sources. Furthermore, the time of dissimilar sources is synchronized by pulse-mode operation. Therefore, although these results highlight the feasibility of quantum interference from hybrid quantum systems, their complexity and inefficiency limit the practicality of the combined system.

Warm atomic ensembles are ideal platforms compatible with QDs. For example, the 780 nm transition of Rb atoms [33] and the 917 nm transition of Cs atoms [34] spectrally align with GaAs/AlGaAs QDs [15] and InAs/GaAs QDs [18,35], respectively. Moreover, warm atomic ensembles exhibit a spectral linewidth ranging from hundreds of MHz to several GHz [36,37], comparable to that of QDs. In particular, a photonic quantum light source using an atomic vapor cell has been actively demonstrated as an excellent candidate due to its compactness and operational simplicity for narrow-band quantum sources [33,34,38]. Notably, bright and resilient spontaneous four-wave mixing (SFWM) photon pairs generated from warm atomic vapor cells have successfully served as crucial resources for experimental realization in quantum memory, quantum repeaters, and long-distance quantum networks [3,23,39,40]. However, direct quantum interference between single photons from these two quantum sources has yet to be demonstrated.

Here, we demonstrate two-photon interference between single photons emitted from a single QD and a warm atomic ensemble. Specifically, $^{133}$Cs atoms generate heralded single photons at well-defined 917 nm, which lies within the emission range of InAs/GaAs QDs. Employing fine spectral tuning of the QD emission enables spectral matching between two quantum sources with a spectral overlap of around 92%. Utilizing the thin vapor cell and nano-cavity integration, both quantum systems experimentally produce single photons efficiently at detection rates approaching the MHz range. As a result, two independent quantum systems efficiently generate quantum interference without tight spectral filtering and time synchronization. Our result provides fundamental insight into how quantum coherence and indistinguishability can be preserved across different physical platforms. Furthermore, our approach establishes a critical step toward interfacing two different quantum systems and paves the way for scalable hybrid quantum networks that support interference and interaction between dissimilar quantum platforms.

# Results

Quantum interference between dissimilar sources requires frequency matching as well as similar coherence times so that the emitted photons are indistinguishable. Before we conduct quantum interference between photons from a warm atomic ensemble and a single QD, we characterize the optical properties of each single-photon source.

### Heralded single-photon generation via SFWM in a warm $^{133}$Cs ensemble

A warm $^{133}$Cs ensemble generates heralded single photons via the SFWM process from its $6S_{1/2} - 6P_{3/2} - 6D_{5/2}$ transition [34]. Figure 1a shows an optical setup of heralded single-photon generation using a mm-scale warm $^{133}$Cs vapor cell [34] (Inset of Fig.1a). Two counter-propagating continuous wave (CW) lasers, called pump and coupling, through warm $^{133}$Cs ensemble (Fig. 1a) produced correlated signal (917.48 nm) and idler (852.36 nm) photon pairs (Fig. 2a). The signal photon count rate was controlled by varying the pump laser power and was tuned between 0.11 MHz and 0.88 MHz during the experiment. In photon-pair generation experiments using an atomic vapor cell, reabsorption effects set an optimal range for the optical depth [37,41]. Employing a thin vapor cell improves the spatial overlap between the pump and the generated modes, thereby suppressing reabsorption losses. As a result, the source exhibits higher brightness and an improved signal-to-noise ratio at a higher optical depth [34]. Under these optimized conditions, the heralding efficiency was significantly improved, reaching as high as 22%. The detailed experimental condition is explained in Methods. The spectrum of the signal photons is shown as a blue solid line in Fig. 2c(left), but the measured spectral linewidth of the heralded single photon from the warm $^{133}$Cs ensemble is limited by the spectrometer's resolution (7 GHz). By using a scanning Fabry-Perot interferometer, as shown in the blue solid line in Fig. 2c(right), we could directly measure the spectrum of heralded single photons, corresponding to the biphoton spectral wavefunction of our photon pairs.

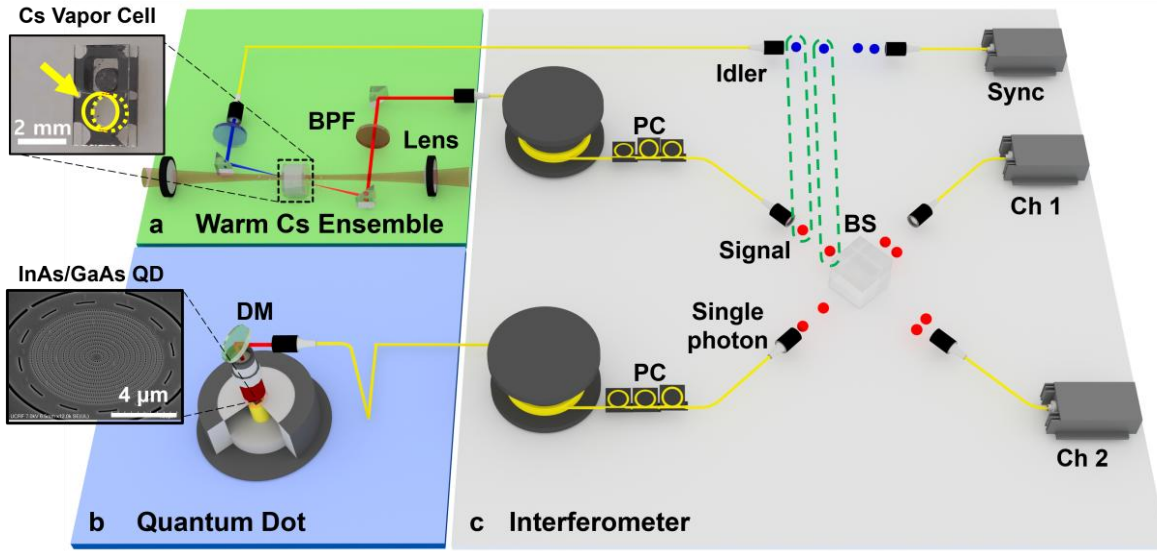

**Fig. 1. Experimental setup of the two-photon interference between two different quantum emitters. a**, Generation of heralded single photons from a warm $^{133}$Cs ensemble (Green regime). The inset shows an image of a cell filled with a warm $^{133}$Cs atomic ensemble. **b**, Deterministic generation of single photons from a cavity-coupled QD, cooled by a He-closed cryostat (Blue regime). The inset shows a scanning electron microscopy image of the fabricated hole-based circular Bragg grating cavity. **c,** Setup for two-photon interference where a signal-idler photon-pair from an atomic cell and a single photon from a QD interfere (Gray regime). BPF: band-pass filter; DM: dichroic mirror; PC: polarization controller; BS: beamsplitter.

**Single-photon generation using QD**

Self-assembled InAs/GaAs QDs are among the most promising platforms for generating high-performance single photons, and their emissions lie around the $6P_{3/2} - 6D_{5/2}$ transition of the warm $^{133}$Cs ensemble. However, unlike atomic transitions, the inhomogeneous nature of QDs leads to their emission over a wide spectral distribution. To achieve spectrally resonant emission from a QD, we fabricated a nanophotonic cavity designed at 917 nm (see Methods). The hole-based circular Bragg grating low-$Q$ cavity (Inset of Fig. 1b) can be coupled with QDs over a broad spectral range, two orders of magnitude broader than single QDs' linewidth [18,19,22]. Finite-difference time-domain simulation shows that single photons emitted from the cavity-coupled QDs exhibit high vertical directionality, improving coupling efficiency with a single-mode optical fiber (NA 0.13) up to 27%.

We first found a cavity-coupled bright QD near the atomic transition of 917.48 nm. The QD was cooled to cryogenic temperature and optically excited using a CW laser at 780 nm (Fig. 1b). The experimental single photon count rate achieves 0.44 MHz. At the lowest temperature of 7.5 K, the QD exhibits a spectral detuning of 0.03 nm relative to the heralded signal photons, along with a spectrometer-limited linewidth (Fig. 2c(left)). To compensate for the small spectral mismatch between the QD emission and the heralded signal photons from the warm $^{133}$Cs ensemble, we red-shifted the QD emission by increasing the sample temperature, achieving resonance at 12.4 K, as shown in Fig. 2b and 2c(left).

To quantify the spectral similarity between two sources beyond the spectrometer's resolution,

we compare their high-resolution spectra measured using the scanning Fabry-Perot interferometer with a transmission window of less than 55 MHz. Under spectrally resonant conditions, we calculate the spectral overlap $A$ as

$$A(S_{QD}, S_s) = \int S_{QD}(\omega) \cdot S_s(\omega) d\omega / \sqrt{\int |S_{QD}(\omega)|^2 d\omega \cdot \int |S_s(\omega)|^2 d\omega} \quad (1)$$

where $S_{QD}$ is the QD emission spectrum and $S_s$ is the spectrum of signal photons heralded by coincidence with the idler photons. The red solid line in Fig. 2c(right) represents a Lorentzian fit ($S_{QD}$) to the QD spectrum at 12.4 K, with a full width at half maximum (FWHM) of $2.17 \pm 0.17$ GHz. The heralded signal photon spectrum ($S_s$, blue solid line in Fig. 2c(right) from the warm $^{133}$Cs ensemble exhibits a more complex profile due to both spectral deformation caused by hyperfine structures of $6P_{3/2}$ and $6D_{5/2}$ levels and photon reabsorption [42]. From the measured $S_{QD}$ and $S_s$, we determine spectral overlap of $A = 0.92$, indicating a high degree of spectral identity between the two dissimilar quantum light sources. In previous studies [27,28,31,32], etalon filters with narrow transmission bandwidths were inevitable due to the large mismatch in the spectral profiles between the sources. This causes a significant loss in their emission and lowers the rate of quantum interference. In contrast, the warm $^{133}$Cs ensemble and the single QD experimentally inherently generate single photons with high brightness of near MHz single photon count rates and high spectral similarity. This intrinsic compatibility ensures two-photon interference without photon losses, representing a key advantage of a hybrid quantum architecture based on a warm atomic ensemble and semiconductor QDs.

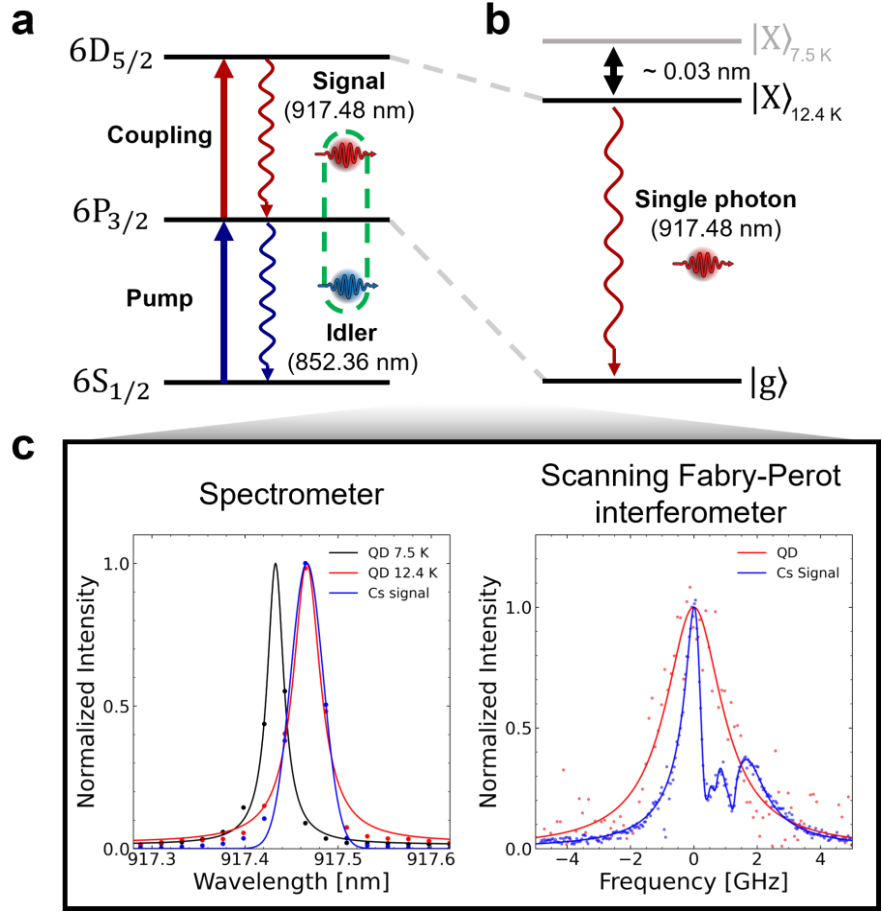

**Fig. 2. Spectral characteristics of two distinct quantum systems. a**, Description of the ladder-type energy level of warm $^{133}$Cs atoms. Red and blue solid arrows indicate the counter-propagating coupling and pump CW lasers. Red and blue curved arrows indicate the correlated signal and idler photons via the SFWM process. **b**, Energy level of a single QD before and after thermal tuning. After thermal tuning, the transition energy becomes resonant with the signal photon of $^{133}$Cs atoms at 12.4 K. **c**, Comparison of spectra of the photons from the warm $^{133}$Cs ensemble and the QD by a spectrometer (left) and high-resolution spectra by a scanning Fabry-Perot interferometer (right). Black and red solid lines represent single photons from the cavity-coupled QD at 7.5 and 12.4 K, respectively. Blue solid lines are the spectra of the signal photons from the warm $^{133}$Cs ensemble.

**Photon correlations and coherence times**

We then perform single-photon characterization of each source by correlation measurements. First, we measured the cross-correlation, $g^{(2)}_{s,i}(\tau)$, between the signal and idler photons from the warm $^{133}$Cs ensemble. Inset of Fig. 3a shows the result with the fitted Gaussian function of $128 \pm 4$ ps FWHM (red solid line in Fig. 3a), indicating the effective heralding time window between the signal and idler photons. The heralded signal photons effectively serve as single photons within this heralding time window, corresponding to the coherence time of the single photons. The cross-correlation amplitude at zero-time delay exceeds $g^{(2)}_{i,s}(0) > 1{,}200$, indicating a strong temporal correlation between signal and idler photons (Inset in Fig.3a).

Owing to the use of a thin vapor cell, the heralding efficiency is directly estimated to be approximately 18% within the experimental heralding time window of 320 ps at individual signal and idler photon count rates of 0.88 MHz. The achieved heralding efficiency is higher than that in previous studies [34,41]. Then, using the idler photons as heralding triggers within the experimental heralding time window of 80 ps, we measured the conditional second-order coincidence of the heralded signal photons. After normalizing the conditional coincidence result (Supplementary Note 1), we get the normalized conditional second-order correlation ($g^{(2)}_{i,s,s}(\tau)$) of heralded signal photons from the warm $^{133}$Cs ensemble (Fig. 3a). The antibunching feature is $g^{(2)}_{i,s,s}(0) = 0.002$, representing the high single-photon purity of the heralded signal photons. The oscillatory features, as shown in the complex frequency spectrum of the signal photon in Figure 2c (right), arise from reabsorption of idler photons, which modifies the spectrum of the heralded signal photons [38,42], and from the quantum beating between the hyperfine levels of $6P_{3/2}$ and $6D_{5/2}$ states in $^{133}$Cs atoms. To characterize the coherence property of the heralded signal photons, we perform two-photon interference between two heralded signal photons generated in opposite directions from the same $^{133}$Cs vapor cell, using the fourth-order coincidence of two idler photons as heralding triggers and two interfering signal photons. To control the indistinguishability of the heralded signal photon, we adjusted the polarizations of the signal photons before passing through the BS. With an 80 ps heralding window applied to the signal-idler photon pairs, we observed a Hong-Ou-Mandel (HOM) dip of $g^{(2)}_{HOM,i,i,s,s}(0) = 0.07 \pm 0.15$, corresponding to the HOM visibility of 93%, as shown in Fig. 3b. After deconvolving the temporal resolution of our system, the HOM visibility is increased to $1 \pm 0.06$. Given the coherence time ($95 \pm 21$ ps) of signal photons from the warm $^{133}$Cs ensemble, the two-photon interference feature can be recorded using time-resolving detection without adjusting the temporal delay. The large interference pattern contrast shows the high indistinguishability of the heralded single photons from the warm $^{133}$Cs ensemble.

Next, we characterize the single-photon properties of the single QD. In contrast to the heralded single photons from the warm $^{133}$Cs ensemble, the single photons from the single QD do not require a heralding process. It shows antibunching behavior via a direct Hanbury Brown and Twiss (HBT) experiment. Figure 3c shows the measured second-order correlation curve of the single QD at 12.4 K. To account for a temporal resolution (104 ps) of the system, we convolved the fitting function,

$$g^{(2)}_{QD}(\tau) = 1 - \left(1 - g^{(2)}_{QD}(0)\right)\exp[-|\tau|/\tau_{QD}] \quad (2)$$

with a convolution of timing jitter of two superconducting nanowire single-photon detectors (SNSPDs) and a time-correlated single photon counting (TCSPC), where $\tau_{QD} = 1.01 \pm 0.02$ ns is the lifetime of a QD. As a result of the low-$Q$ cavity coupling, the single-photon count rate achieved 0.44 MHz, and the pronounced antibunching profile with $g^{(2)}_{QD}(0) = 0.01 \pm 0.01$ confirms the generation of high-purity single photons from the single QD. We subsequently performed HOM interferences on the single photons from the single QD to assess their indistinguishability and coherence time. Temporally separated single photons were interfered with at a 50:50 BS within an asymmetric Mach-Zehnder interferometer with an optical delay of 4 m (= 19.6 ns) in one of the interferometer arms. Figure 3d displays HOM results for the cases: two orthogonally (black, distinguishable) and parallelly (red, indistinguishable) polarized photons after convolving the temporal resolution of our system. A HOM interference dip below 0.5 appears when the input photons are indistinguishable ($g^{(2)}_{HOM,\parallel}(0) < 0.5$). From the calculation of $V_{QD}(\tau) = \left(g^{(2)}_{HOM,\perp}(\tau) - g^{(2)}_{HOM,\parallel}(\tau)\right)/g^{(2)}_{HOM,\perp}(\tau)$, we obtained uncorrected single-photon visibility of $V_{QD}(0) = 0.62 \pm 0.07$ at 12.4 K. After deconvolving the temporal resolution of our system, we finally obtained $V_{QD}(0) = 1 \pm 0.10$, representing a high degree of single-photon indistinguishability with a coherence time of $\tau_{QD,c} = 129 \pm 18$ ps. We could control the central frequency of the single photon from the QD by engineering the QD sample temperature with high single-photon indistinguishability. Through these characterizations, we compared two different types of single-photon sources and confirmed physical similarities.

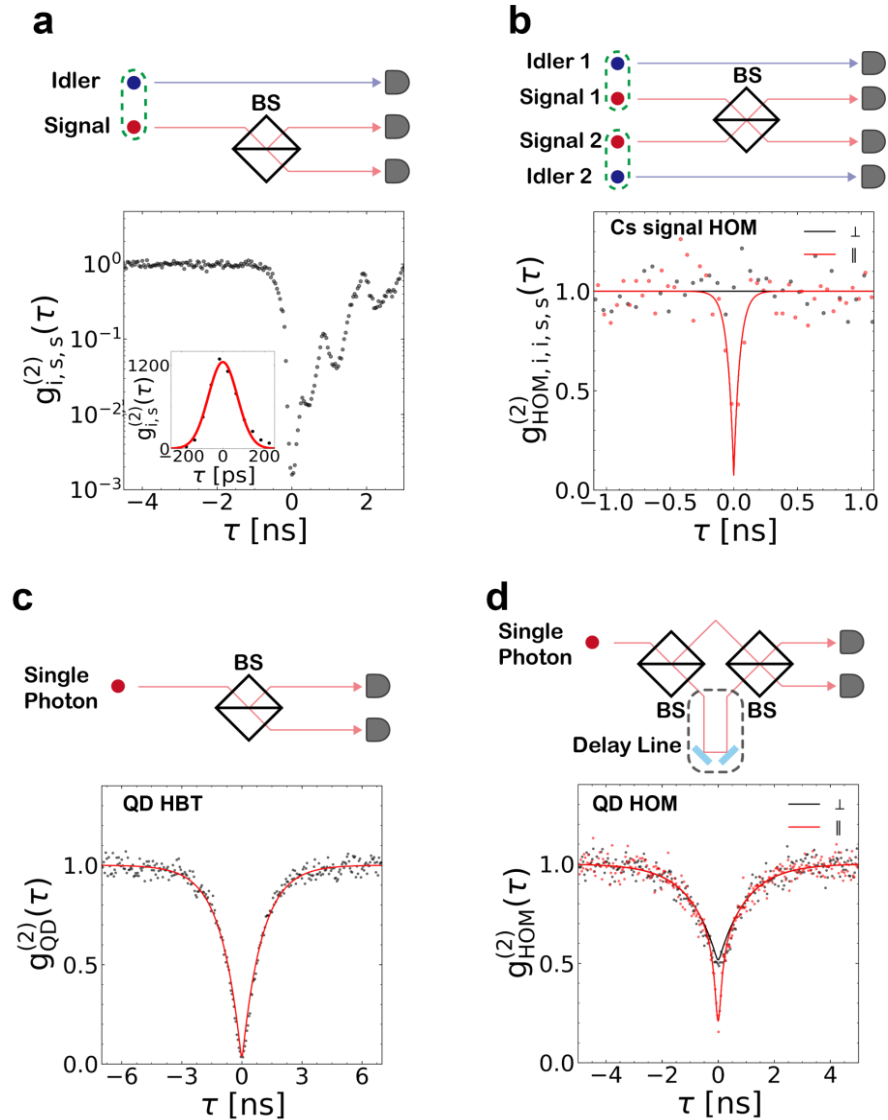

**Fig. 3. Characterizations of single photons from the warm $^{133}$Cs ensemble and the QD with their coherence properties. a**, Normalized second-order correlation curve for the signal photons from the warm $^{133}$Cs ensemble by using a three-fold coincidence setup. The inset is the cross-correlation curve between the signal and idler photons, exhibiting strong coincidence counts with the Gaussian FWHM of $128 \pm 4$ ps. **b**, HOM interference curves for the signal

photons using a four-fold coincidence setup. The black and red lines represent interference patterns for orthogonally (distinguishable) and parallelly polarized (indistinguishable) two input signal photons, respectively. **c**, HBT measurement of the single photon from the cavity-coupled QD at 12.4 K. **d**, HOM interference measurement of single photons from the QD. The black (red) line represents the interference pattern for the orthogonally (parallelly) polarized cases. Schematics above the graphs represent brief experimental schemes for correlation measurements. A 4 m delay line was inserted in one arm to interfere with two temporally separated single photons.

**Two-photon interference between autonomous dissimilar quantum light sources**

Finally, we demonstrate two-photon interference between single photons from the QD and heralded signal photons from the warm $^{133}$Cs ensemble under continuous excitation, as described in Fig. 1c. For the two-photon interference between deterministic and heralded single photons from two different quantum light sources, we sent single photons from the QD into one input port of the BS and sent signal photons into the other port. The correlated idler photons served as the heralding trigger, enabling the three-fold coincidence measurements. To investigate both distinguishable and indistinguishable cases, the polarization of the input photons was adjusted accordingly. The output photons from the BS were then detected by two SNSPDs, labeled Ch1 and Ch2, while the idler photons were separately recorded by the Sync channel. By selecting conditional coincidence events in which the idler photons were detected within the 80 ps heralding time window at Sync, we measured the three-fold coincidence, $C_{i,s,QD}(\tau)$, between Ch1 and Ch2, ensuring that the single photons interfered only with the heralded signal photons. The three-fold coincidence ($C_{i,s,QD}(\tau)$) can be represented as follows:

$$C_{i,s,QD}(\tau) = \left(1 + \left(C_{i,s,QD}(0) - 1\right)\exp[-2|\tau|/\tau_{hs}]\right) \times \left(1 - I_{id}\exp[-2|\tau|/\tau_{i,s,QD}]\right) \quad (3)$$

where $\tau_{hs}$ is the coherence time of heralded signal photons, $\tau_{i,s,QD}$ is the coherence time of the two-photon interference, and $I_{id}$ is the degree of indistinguishability between interfered signal photons from the warm $^{133}$Cs ensemble and single photons from the QD. If the interfered photons were distinguishable (indistinguishable), $I_{id}$ is 0 (1). HOM visibility is defined as $V_{i,s,QD}(\tau) = \left[C_{i,s,QD}^{dis}(\tau) - C_{i,s,QD}^{indis}(\tau)\right]/C_{i,s,QD}^{dis}(\tau)$, where $C_{i,s,QD}^{dis(indis)}(\tau)$ is the three-fold coincidence for distinguishable (indistinguishable) photon inputs. For comparison, we first performed two-photon interferences when the QD and the warm $^{133}$Cs ensemble are spectrally detuned ($\delta = 0.03$ nm). In Fig. 4a, the results show that there is no visible change in their coincidence events between orthogonal and parallel polarized photon cases, which is expected since they are spectrally distinguishable. Then, we performed two-photon interference for the case of resonant two sources after spectral tuning of the QD. The two-photon interference suppresses coincidence events at zero-time delay when two single photons are parallelly polarized, and we observe a large interference contrast (Fig. 4b). This comparison confirms that spectral indistinguishability is essential for two-photon interference between photons from two different quantum light sources. We note that the photon count ratio ($R_{s/QD}$) of unheralded signal photons from the warm $^{133}$Cs ensemble to single photons from the QD determines the amount of suppression and, consequently, the visibility of two-photon interference (See Supplementary Note 2). Hence, we investigate two-photon interference according to $R_{s/QD}$. To vary $R_{s/QD}$, we adjust the count rate of unheralded signal photons from the warm $^{133}$Cs ensemble while fixing the count rate of single photons from the QD. To quantify the degree of successive two-photon interference, we calculated the visibility and fitted the visibility curves using a fitting function:

$$V_{i,s,QD}(\tau) = V_{i,s,QD}(0)\exp[-2|t|/\tau_{i,s,QD}] \tag{4}$$

where $\tau_{i,s,QD}$ is the coherence time of the two-photon interference. To account for the finite system temporal resolution of 124 ps (3 SNSPD detectors and TCSPC), we convoluted the coincidence curve with our system's temporal response and corrected the interference visibilities. Figure 4c plots the measured and corrected visibility as a function of $R_{s/QD}$. The corrected visibility increases as $R_{s/QD}$ decreases. At a low ratio of $R_{s/QD} = 0.25$, the polarization-dependent interference pattern contrast is enhanced, as shown in Fig. 4b, and the deconvoluted visibility rises to $V_{i,s,QD}(0) = 0.65 \pm 0.14$.

To analyze how $R_{s/QD}$ affects two-photon interference visibility, we developed a theoretical model under realistic lossy conditions and performed numerical calculations. Analyzing two-photon interference between the single photons from the QD and heralded signal photons from the warm $^{133}$Cs ensemble involves the three-fold coincidence among Ch1 and Ch2 (interfered photons) and Sync (the idler photons), defined as:

$$C_{i,s,QD}(0) = \langle \hat{a}_i^\dagger \hat{a}_{Ch1}^\dagger \hat{a}_{Ch2}^\dagger \hat{a}_{Ch2} \hat{a}_{Ch1} \hat{a}_i \rangle \tag{5}$$

where $\hat{a}_{Ch1}^\dagger$, $\hat{a}_{Ch2}^\dagger$, and $\hat{a}_i^\dagger$ are creation operators for photons detected at Ch1, Ch2, and Sync, respectively. Following refs [43], the three-fold coincidence can be expressed as

$$C_{i,s,QD}(0) = [\langle \hat{a}_i^\dagger \hat{a}_{QD}^\dagger \hat{a}_{QD}^\dagger \hat{a}_{QD} \hat{a}_{QD} \hat{a}_i \rangle + \langle \hat{a}_i^\dagger \hat{a}_s^\dagger \hat{a}_s^\dagger \hat{a}_s \hat{a}_s \hat{a}_i \rangle + 2\langle \hat{a}_{QD}^\dagger \hat{a}_{QD} \rangle \langle \hat{a}_i^\dagger \hat{a}_s^\dagger \hat{a}_s \hat{a}_i \rangle (1 - I_{id})] \tag{6}$$

where $\hat{a}_{QD}^\dagger$ and $\hat{a}_s^\dagger$ are creation operators for the single photons from the QD and the signal photons from the warm $^{133}$Cs ensemble before entering the BS, and $I_{id}$ represents the degree of indistinguishability between the single photons from the QD and heralded signal photons from the warm $^{133}$Cs ensemble (Supplementary Note 2). Taking into account our experimental system efficiencies of $\eta_s = 0.37$ and $\eta_i = 0.57$ for the signal and idler photons, we numerically calculated the two-photon interference visibility (Supplementary Notes 2 and 3). Figure 4d shows the calculated interference visibility as a function of system efficiency for the single photons generated by the QD ($\mu$) and a mean photon number of unheralded signal photons from the $^{133}$Cs ensemble ($\bar{n}$). At the high mean photon number of unheralded signal photons, the heralded signal photons no longer serve as heralded single photons due to substantial multi-photon components, leading to interference visibility reduction. In contrast, the high system efficiency of our warm $^{133}$Cs ensemble suppresses uncorrelated background, thereby enabling more robust two-photon interference than systems of lower efficiency at the same mean photon number ($\bar{n}$) conditions. High interference visibility also requires low $R_{s/QD}$. In the QD system, high $\mu$ directly increases the single photon count rate, thereby lowering $R_{s/QD}$ and enhancing interference visibility at fixed $\bar{n}$ of the warm $^{133}$Cs ensemble system. Therefore, to achieve fast and high two-photon interference visibility in actual experiments, it is beneficial to use heralded signal photons with a small mean photon number and to ensure high system efficiency in both systems.

Considering our experimental conditions, we compare the experimental results with theoretical models as a function of $R_{s/QD}$ (Fig. 4c). Since the count rate of the unheralded signal photons directly determines $\bar{n}$, and $R_{s/QD}$ is proportional to $\bar{n}$, $R_{s/QD}$ serves as a key parameter of the interference visibility. The red curve represents the expected interference visibility, including the finite temporal resolution of 124 ps and the coherence time ($\tau_{i,s,QD}$) of 119 ps. The fact that the coherence time of single photons is comparable to the system's temporal resolution becomes a major factor in limiting the visibility. The blue curve plots the theoretical model, assuming infinitesimal temporal resolution, which enhances the interference visibility. All experimental data closely follow the theoretical predictions with $\mu = 0.019$ (Supplementary Note 3).

This result shows the first experimental two-photon interference between two different types of quantum light sources without spectral filtering of single photons from each source and time synchronization, such as a pulse operation. Our theoretical model well describes the two-photon interference as a function of the count rate ratio. At $R_{s/QD} = 0.25$ condition, we obtained the interference visibility of 0.65. With a tenfold improvement in our experimental system efficiency for the QD ($\mu \approx 0.2$), which is comparable to the warm $^{133}$Cs ensemble system, the two-photon interference visibility is expected to exceed 0.95 by achieving much lower $R_{s/QD}$. Such efficiency enhancement can be achieved by employing various quantum platforms [22,44] to minimize losses in the experimental system for the QD.

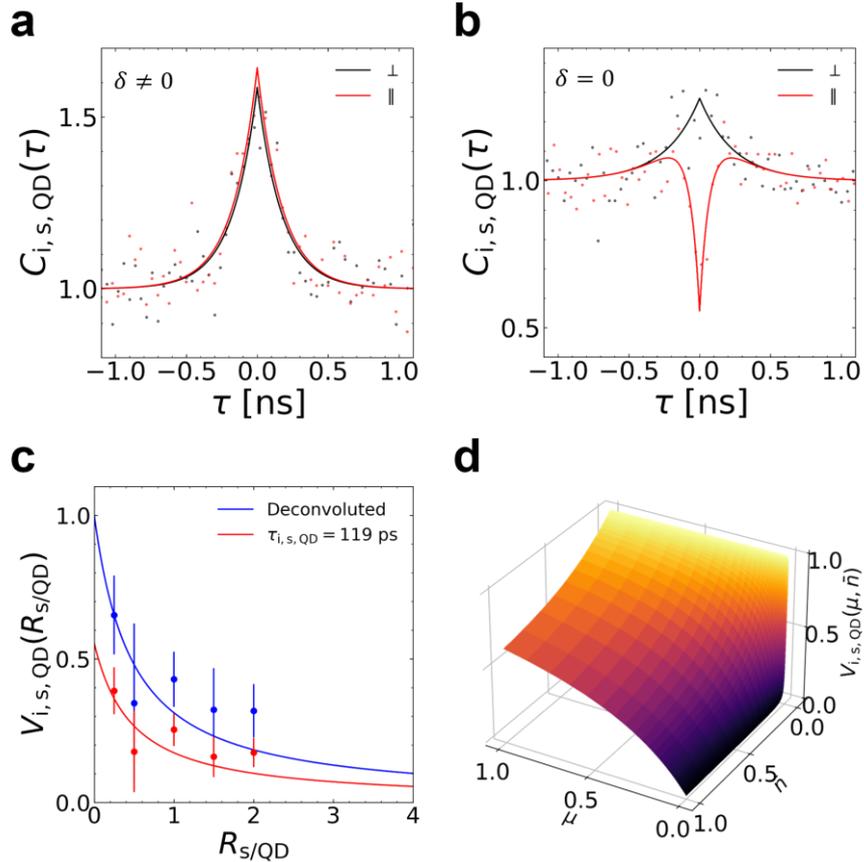

**Fig. 4. Two-photon interference between the single photons and the heralded signal photons. a**, Two-photon interference measurement when the QD is spectrally detuned ($\delta \neq 0$) from the warm $^{133}$Cs ensemble transition. The black and red data represent distinguishable (orthogonal) and indistinguishable (parallel) in polarization basis, respectively. The corresponding fitted bunching amplitudes are $C_{i,s,QD}^{dis}(0) = 1.59 \pm 0.05$ and $C_{i,s,QD}^{indis}(0) = 1.64 \pm 0.03$. **b**, Two-photon interference measurements for resonant ($\delta = 0$) between the two single photons from the QDs and the warm $^{133}$Cs ensemble when $R_{s/QD} = 0.25$. The coincidence amplitudes are $C_{i,s,QD}^{dis}(0) = 1.28 \pm 0.04$ and $C_{i,s,QD}^{indis}(0) = 0.56 \pm 0.07$. **c**, Experimentally measured (points) and theoretically calculated (solid lines) two-photon interference visibilities. The red and blue data represent measured and calculated results with and without accounting for the system resolution. **d**, Analytical model for the interference visibility ($V_{i,s,QD}(0)$) as a function of mean photon number ($\bar{n}$) of unheralded signal photons and the system efficiency ($\mu$) of the QD.

## Discussion

A quantum network requires interfacing stationary and photonic qubits, with all qubits across distributed quantum nodes operating at identical wavelengths. In this study, we demonstrate quantum interference between two photons generated from a thin atomic cell and a cavity-coupled QD. Both platforms are capable of emitting bright, indistinguishable single photons with inherently similar spectral and temporal profiles, which is essential for efficient interfacing and for building modular hybrid quantum networks. This result marks a significant advance in the development of hybrid quantum architectures, where photons from distinct quantum systems must interfere coherently.

In our hybrid system, interference is directly observed from the single-photon emission of each source, distinguishing from previous hybrid demonstrations that relied on external filtering to overcome spectral mismatches. At present, the non-ideal visibility primarily originates from the limited coherence time and system inefficiencies. The QD's coherence time could be extended through (quasi-)resonant excitation techniques [18,45]. Also, instead of thermal tuning of QDs, applying an external field [46] could expand the tuning range without introducing additional dephasing.

Another importance of the work is that QDs and atomic cells each have distinct advantages for generating and storing single photons. Also, atoms can provide a global frequency reference for ensuring spectral identity between remotely separated QDs. The optical compatibility between these heterogeneous platforms, therefore, not only bridges the gap between photon generation and storage, but also provides a frequency standard, paving the way for long-distance entanglement distribution and the development of quantum repeaters.

Our findings establish that interfacing heterogeneous quantum platforms via indistinguishable single photons. In the future, by implementing Bell-state measurements following two-photon interference, quantum teleportation could be realized [47,48]. Our work offers a scalable and practical route to a modular hybrid quantum system, an essential step toward future quantum network architectures.

# Methods

## InAs/GaAs sample design and fabrication

For the optimization and fabrication of the cavity, we followed our previous work [18]. The hole-base circular Bragg grating cavity was optimized for maximizing the collection efficiency of 907 nm emission using a finite-difference time-domain method. Optimized parameters are radial direction periodicity: $\Lambda = 322$ nm, center disk diameter: $R = 2.6\Lambda$, tangential direction periodicity: $w = 0.43\Lambda$, and hole diameter: $r = 0.30\Lambda$. The optimized Gaussian far-field exhibits directional emission within 5-degree divergence.

Via the molecular beam epitaxy method, self-assembled InAs QDs embedded in 160 nm GaAs were grown. We deposited an additional $Si_3N_4$ layer with a thickness of 95 nm. We made a dry etch mask using electron beam lithography and dry etched $Si_3N_4$ and GaAs layers using reactive ion etching processes. To make the sample air-suspended membrane structure, we selectively etched the AlGaAs sacrificial layer under the GaAs layer using diluted hydrofluoric acid.

## Experimental setup

In one arm, a 1-mm-long cesium vapor cell, heated to 105°C, was simultaneously illuminated by the CW 852 nm pump and 917 nm coupling lasers arranged in a counter-propagating geometry. The lasers were detuned by $+1.35$ GHz (one-photon) and $-40$ MHz (two-photon), respectively. While the coupling laser power was fixed at 15 mW, the pump laser was varied below 0.03 mW to adjust the heralded single-photon generation rate. The phase-matched photon pairs were emitted at specific angles; the signal and idler photons, emerging at approximately a 2-degree angle with respect to the counter-propagating lasers, were filtered with a band-pass filter and polarizer to serve as the heralded single-photon source. Across all experimental settings, the heralding efficiency, defined as the ratio of coincidence counts to idler counts, remained above 9% (18%, 22%) within 80 ps (320 ps, 2 ns) experimental heralding time window.

In the other arm, the InAs QD was cooled down to 7.5 and 12.4 K using a closed-cycle He-flow cryostat (ST-500, Lake Shore) and excited by the CW laser of 780 nm. Then, polarization-filtered QD emission was coupled into one input port of a fiber interferometer. Before the two-photon interference, the single photons were spectrally isolated using a grating-based fiber band-pass filter (WL Photonics) of 50 GHz bandwidth.

At the 50 : 50 beam splitter, the QD photon and the heralded signal photon were superposed, with PCs adjusted to parallel or orthogonal polarization to enable or suppress two-photon interference. Three-fold coincidence counts among the idler and two output channels of the interferometer were recorded to quantify the interference visibility. All photons were detected using SNSPDs optimized near 780 nm, with quantum efficiencies of approximately 70% at 852 nm and 50% at 917 nm, and a timing jitter of ~70 ps FWHM. Time-tagging and correlation analysis were performed with a quTAG-MC (qutools) TCSPC, which provides 50 ps FWHM timing resolution. For the precise spectrum measurement, we used the scanning Fabry-Perot interferometer (SA210-8B, Thorlabs) with a free spectral range of 10 GHz and a finesse of 160, combined with a function generator to adjust the interferometer's transmission frequency.

## Data availability

The data that support the findings of this study are available from the corresponding authors upon request.

## Code availability

The custom codes for simulating the quantum system are available from the corresponding authors upon request.

## Acknowledgements

The authors acknowledge financial support from the National Research Foundation of Korea grant funded by MSIT (RS-2023-00283146, RS-2024-00442762, and NRF-2022R1A2C2003176), the Institute for Information & Communications Technology Planning & Evaluation (IITP) Grant (RS-2024-00338878 and IITP-2025-RS-2020-II201606), and the Institute for Information and Communications Technology Promotion (RS-2024-00396999, IITP-2025-2020-0-01606, and IITP-2022-0-01029). We acknowledge support from the 2025 Research Fund (1.250007.01) of UNIST (Ulsan National Institute of Science & Technology).

## Author contributions

K.-Y. Kim and H. Kim equally contributed to this paper. H. Kim, J. Bea, and G. Ju prepared heralded signal photons using the warm $^{133}$Cs ensemble. S. I. Park and J. D. Song grew the QD sample. K.-Y. Kim designed and fabricated the QD sample. K.-Y. Kim and D. H. Park prepared single photons using the QD. K.-Y. Kim and H. Kim performed the optical experiments and prepared the figures. K.-Y. Kim performed theoretical calculations. K.-Y. Kim, H. Kim, J. Bea, J.-H. Kim, and H. S. Moon analyzed the data and wrote the manuscript with input from all authors. J.-H. Kim and H. S. Moon supervised the project.

## Competing interests

The authors declare that they have no competing interests.

## Additional information

**Supplementary information** is available for this paper.

**Correspondence and requests for materials** should be addressed to Je-Hyung Kim and Han Seb Moon.

**Reprints and permissions information** is available at www.nature.com/reprints.